# Prediction of potential inhibitors for RNA-dependent RNA polymerase of SARS-CoV-2 using comprehensive drug repurposing and molecular docking approach

Md. Sorwer Alam Parvez[1,#], Md. Adnan Karim[2,#], Mahmudul Hasan[3,#], Jomana Jaman[4], Ziaul Karim[5], Tohura Tahsin[1], Md. Nazmul Hasan[2], Mohammad Jakir Hosen[1,*]

[1]Department of Genetic Engineering & Biotechnology, Shahjalal University of Science & Technology, Sylhet-3114, Bangladesh

[2]Department of Genetic Engineering & Biotechnology, Jashore University of Science & Technology, Jashore, Bangladesh

[3]Department of Pharmaceuticals and Industrial Biotechnology, Sylhet Agricultural University, Sylhet-3100, Bangladesh

[4]Dept. of Biotechnology, Bangladesh Agricultural University, Mymensingh, Bangladesh

[5]Dept. of Biology, Chemistry and Pharmacy, Free University of Berlin, Berlin, Germany

[#]Joint first author. Authors contributed equally

*Correspondence
Prof. Dr. Mohammad Jakir Hosen
Dept. Of Genetic Engineering & Biotechnology
Shahjalal University of Science &Technology, Sylhet-3114, Bangladesh
E-mail: jakir-gen@sust.edu

**Abstract**

The pandemic prevalence of COVID-19 has become a very serious global health issue. Scientists all over the world have been heavily invested in the discovery of a drug to combat SARS-CoV-2. It has been found that RNA-dependent RNA Polymerase (RdRp) plays a crucial role in SARS-CoV-2 replication, and thus could be a potential drug target. Here, comprehensive computational approaches including drug repurposing and molecular docking were employed to predict an effective drug candidate targeting RdRp of SARS-CoV-2. This study revealed that Rifabutin, Rifapentine, Fidaxomicin, 7-methyl-guanosine-5'-triphosphate-5'-guanosine and Ivermectin have a potential inhibitory interaction with RdRp of SARS-CoV-2, and could be effective drugs for COVID-19. In addition, virtual screening of the compounds from ZINC database also allowed the prediction of two compounds (ZINC09128258 and ZINC 09883305) with pharmacophore features that interact effectively with RdRp of SARS-CoV-2; indicating their potentiality as effective inhibitors of the enzyme. Furthermore, ADME analysis along with analysis of toxicity was also investigated to check the pharmacokinetics and drug-likeness properties of the two compounds. Comparative structural analysis of protein-inhibitor complexes revealed that positions of the amino acid Y32, K47, Y122, Y129, H133, N138, D140, T141, S709 and N781 are crucial for drug surface hotspot in the RdRp of SARS-CoV-2.

**Keywords:** COVID-19; SARS-CoV-2; RNA-dependent RNA polymerase; Drug

1. **Introduction**

The pandemic Corona Virus Disease 19 (COVID-19) has become a critical, rapidly emerging public health issue for the world. It is caused by the outbreak of Severe Acute Respiratory Syndrome Corona Virus 2 (SARS-CoV-2), the disease characterized by fever, cough, and severe shortness of breathing, nausea, vomiting and diarrhea [1]. As of 10 April 2020, SARS-CoV-2 infection has been reported in 184 countries with 1,536, 979 confirmed cases and 93,425 total deaths [2,3]. Epidemiological data have determined person to person transmission as the route of the rapid outbreak of COVID-19, which has become a major obstruction in combating the virus [4,5].Clinical studies reported that older patients have a higher case of fatality rate (CFR) than the young, and males have a higher CFR than female [6]. Apart from acute respiratory distress, COVID-19 patients have been diagnosed with higher rate of renal impairment, indicating the development of kidney dysfunction[7]. Unfortunately, there are no proven drugs, vaccines or therapies available to fight against COVID -19.

SARS-CoV-2 is an enveloped, positive-sense, single-stranded RNA virus similar to SARS and MERS (Middle East Respiratory Syndrome) coronavirus [8]. Along with structural proteins (like spike glycoprotein and accessory proteins), the viral genome also encodes non-structural proteins, including 3-chymotrypsin-like protease, papain-like protease, helicase and RNA-dependent RNA polymerase (RdRp)[9]. RdRp is an essential enzyme involved in the replication of RNA viruses including SARS-CoV-2. Several anti-viral drugs have been developed targeting this enzyme for treating infections like Hepatitis C, Zika and other coronaviruses[10]. Although not yet extensively explored, some of these drugs also target the SARS-CoV-2 RdRp or its catalyzed polymerization process[11–14]. A recent study suggested that two known anti-viral drugs Remdesivir and Favipiravir which are used for the treatment of variety of RNA virus diseases, by targeting RNA polymerase and RdRp respectively, could successfully inhibit SARS-CoV-2 replication *in vitro*[13–17].Unfortunately, the mechanism of action and the efficacy of these compounds remain unclear.

Computation biology and molecular docking approach have a wide variety of applications in drug discovery, giving novel insights to the screening of potential drugs for the treatment of COVID-19. These approaches also aid the understanding of the protein-ligand interactions as well as in figuring out drug surface hotspot, which are important for the discovery of an effective drug[18,19]. In addition, drug repurposing approach is used for the identification of existing drugs for one disease, for thetreatment of another disease. As the long-term solution, vaccine, will take years to be marketable, effective repurposing of existing drug remains the only alternative way to fight against an emerging disease like COVID-19[20–23]. Thus, in the presented study, drugs with proven anti-viral activity wereanalyzed using molecular docking and pharmacophore modeling technique to target RdRp of SARS-CoV-2. Our findings can open a new avenue to fight against COVID-19.

2. Materials and Methods

2.1 Retrieval of the structure of SARS-CoV-2 RNA Polymerase and Drug Candidates List

The RdRp structure of SARS-CoV-2 was retrieved from the Protein Data Bank (PDB) with PDB ID: 6M71[24]. Furthermore, 44 drug candidates having inhibitory activity against RNA polymerase were selected by comprehensive literature study, and their PDB structures were retrieved from the Drug Bank Database (Supplementary Table 1) [25]. RdRp inhibitor drugs were prioritized; but drugs that are DNA-dependent RNA polymerase inhibitors having the antiviral activity were also included.

2.2 Screening of RNA polymerase inhibitors against the RdRp of SARS-CoV-2

The AutoDock Vina software of molecular docking approaches was employed for the screening of the drugs against RdRp of SARS-CoV-2[26]. At first, the crystal structure of RdRp was retrieved from the PDB, processed by removing water and complex molecules using PyMOL[27]. After preparing the PDB structures of the inhibitors, they were exposed to the RdRp polymerase of SARS-CoV-2 for analyzing the lowest binding energy and interactive amino acids. The grid box parameters were set to size 80Å x 95Å x 95Å(x, y and z) and centre 121.253Å x 121.376Å x 120.149Å (x, y and z). The 2D ligand-protein interaction diagrams were generated by LigPlot+ to find out the involved amino acids with their interactive position in the docked molecule[28]. Discovery Studio and PyMOL were used to visualize and analyze the ligand molecules' interactions with the viral proteins[29]. Additionally, the Protein-Ligand Interaction Profiler (PLIP) was also used to analyze the total number of non-covalent interactions

(hydrogen bonds, water bridges, salt bridges, halogen bonds, hydrophobic interactions, π-stacking, π-cation interactions and metal complexes) in protein-ligand complexes[30].

## 2.3 Structural Insights of Drug Surface Hotspot in the RdRp of SARS-CoV-2

LigPlot+, Discovery Studio and PyMOL were used to figure out the drug surface hotspot from the docked structures of RdRp with the top-most polymerase inhibitors. Remdesivir and Favipiravir were used as positive control, as they were reported to be effective for COVID-19 by several recent studies[13,14].

## 2.4 Pharmacophore Modelling and Virtual Screening of ZINC Database

PharmaGIST was used for the modeling of the pharmacophore features that are essential for the interaction of RNA polymerase inhibitors with RdRd of SARS-CoV-2. In this study, the top-most inhibitors along with Remdisivir were used for the pharmacophore modeling [31,32]. ZINCPharmer was used to import the generated pharmacophore from PharmaGIST for the virtual screening of the novel compounds from the ZINC database [33,34]. These novel compounds were further used for the screening of new inhibitors of RdRp of SARS-CoV-2. The validity of the screened compounds was checked by Molecular docking approaches.

## 2.5 Drug Likeness Properties Analysis of the Screened Compounds from ZINC Database

SwissADME server was used to assess the Absorption, Distribution, Metabolism and Excretion (ADME) properties of the compounds screened from ZINC database [35]. This server is well known for successfully evaluating the pharmacokinetics, drug-likeness and medicinal chemistry friendliness of potential drug candidates. In this study, the physicochemical parameters (Formula Molecular weight, Molar Refractivity, TPSA), lipophilicity (Log Po/w (iLOGP), Log Po/w (XLOGP3), Log Po/w (WLOGP), Log Po/w (MLOGP), Log Po/w, (SILICOS-IT), Consensus Log Po/w), pharmacokinetic parameters (CYP1A2 inhibitor, CYP2C19 inhibitor, CYP2C9 inhibitor, CYP2D6 inhibitor, CYP3A4 inhibitor, Log Kp; skin permeation) and water solubility (Log S: SILICOS-IT, Solubility) were checked for the screened compounds, also considering all default parameters[36,37]. Additionally, OSIRIS Property Explorer and admetSAR were used to investigate the undesired effects of these compounds like mutagenicity, tumerogenecity and toxicity[38–41]. No ADME or toxicity analysis was required for the top selected drugs since the drugs were previously tested for FDA approval.

## 3. RESULTS

### 3.1 Screening of RdRp inhibitors against the RdRp of SARS-CoV-2

Molecular docking of all the RNA polymerase inhibitors against RdRp of SARS-CoV-2 (PDB ID: 6M71) revealed Rifabutin, Rifapentine, Fidaxomicin, 7-methyl-guanosine-5'-triphosphate-5'-guanosineand Ivermectin chronologically as the top-most effective RdRp inhibitorsof SARS-CoV-2 with highest binding affinity and lowest free energy (Table 1 and Fig. 1). Remarkably, all these top listed inhibitors showed lower binding energy compared to the positive control Remdesivir and Favipiravir. The whole molecular docking results were included in the Supplementary Table 2.

### 3.2 Structural Insights of Drug Surface Hotspot in the RdRp of SARS-CoV-2

The molecular docking pattern and involved amino acid residues with their respective position were further analyzed to reveal the common interactive sites of RdRp in the SARS-CoV-2. Thus, the binding pattern of five most effective compounds along with Remdesivir and Favipiravir, were analyzed to observe the common drug surface hotspot. It was found that the amino acid Y32, K47, Y129, H133, N138, C139, T141 and S709 in RdRp were involved in the interaction with Rifabutin. The amino acid position of Y32, K47, Y129, H133 and S709 along with D140 and N781 were also found crucial for the RdRp of SARS-CoV-2 to interact with Rifapentine, Fidaxomicin, 7-methyl-guanosine-5'-triphosphate-5'-guanosine and Ivermectin (Table 1 and Fig. 2& 3). Most surprisingly, these amino acid residues were also found to be involved in the interaction of Remdesivir (K47, Y129, A130, H133, F134, D135, N138, C139, T141, S709, T710, D711, Q773 and N781) and Favipiravir (Y129, H133, S709, K780 and N781).

### 3.3 Pharmacophore Modeling and Screening of ZINC Database

The top listed RdRp inhibitors along with Remdesivir were further used for the pharmacophore modeling and screening of ZINC database. The pharmacophore modeling (predicted by the

PharmaGIST) revealed 6 spatial features (Aromatic-1, Hydrophobic-1 and Acceptors-4) (Figure 3). This pharmacophore model was imported in the ZINCPharmer for the screening of the ZINC Database, which revealed four different hits (ZINC09128258, ZINC09883305, ZINC09883308 and ZINC11286235). In addition, molecular docking analysis revealed two compounds (ZINC09128258 and ZINC09883305) that could also act as the inhibitors of RdRp of SARS-CoV-2 as they showed similar binding pattern as the top drugs and controls (Table 2 and Fig. 5 & 6). However, these two compounds showed lower binding affinity than the top selected RdRp inhibitors. Additionally, higher number of non-covalent interactions was found for ZINC09883305 (9 non-covalent interactions).

### 3.4 Drug Likeness Properties Analysis of the Screened Compounds from ZINC 15 Database

The physico-chemistry, pharmacokinetics, medicinal chemistry friendliness and toxicity of these two screened compounds (ZINC09128258 and ZINC09883305) from ZINC database were analyzed by SwissADME, ADMETsar and OSIRIS Property Explorer. The physiochemical parameters, lipophilicity and water solubility of these compounds are described in Table 3. Water solubility was also studied in this study and found that all the compounds were moderately soluble. Other important properties such as Molecular weight (MW), molecular refractivity (MR) and topological polar surface area (TPSA), which are very useful for the estimation of ADME properties were also included in this study. Remarkably, none of the screened compounds showed any undesired effects such as mutagenicity, tumorigenicity, irritating and reproductive effects. However, these compounds showed CYP450 enzymes inhibition effects except CYP1A2. Lastly, BOILED-Egg model was employed to calculate the Blood-brain barrier (BBB) permeation that revealed no BBB permeate in the studied compounds[42].

### 4. Discussion

At present, COVID-19 is a global challenge for the scientific communities as its pandemic attitude is dangerously affecting millions of people and taking thousands of lives everyday. But to date, no satisfactory breakthrough has been made in the treatment of Covid-19[43–46]. Several attempts have been made to treat this disease but these drug candidates remain questionable owing to low efficacy[47]. Computational approaches along with drug repurposing methods could be an effective approach to this COVID-19 challenge. In this study, several polymerase inhibitors targeting RdRp of SARS-CoV-19were studied, as RdRp had already shown to be an effective anti-viral drug target for various viral pathogens such as Hepatitis C Virus, HIV, Zika virus etc[48,49]. Here, drug repurposing along with molecular docking was employed for the screening and analysis of the drug candidates against RdRp of SARS-CoV-2. Moreover, the common drug surface hotspot was studied along with modeling of pharmacophore which is very important for drug discovery.  Additionally, novel compounds from ZINC database were screened out which could be developed as a new drug to treat COVID-19.

RdRp plays indispensable roles in the life cycle of RNA viruses. RNA viruses initiate RNA synthesis by virus polymerase utilizing primer-independent and primer-dependent mechanism. Moreover, RdRp based RNA synthesis doesn't occur in the mammalian cells offering an opportunity to design drugs specifically acting against RNA viruses. Additionally, the protein structure of RdRp in RNA viruses is found to be remarkably conserved. Various antiviral drugs have been developed targeting this enzyme for the treatment of infections caused by RNA viruses and they are working effectively[10,50].Therefore, this present study aimed to identify potential drugs targeting this enzyme for the treatment of COVID-19.

Molecular docking analysis revealed that Rifabutin, Rifapentine, Fidaxomicin, 7-methyl-guanosine-5'-triphosphate-5'-guanosine and Ivermectin could be potential RdRp inhibitors of SARS-CoV-2. Notably, Ivermectin has already been reported as showing inhibitor property in SARS-CoV-19 replications *in vitro*[51]. Rifabutin and Rifabutin have also been reported for their anti-viral activity against HIV and vaccinia virus respectively[52,53]. Moreover, the vast number of non-covalent interactions between these screened compounds with RdPd suggests that the protein-inhibitor complexes are very stable. In addition to these drugs, Remdesivir along with Favipiravir which has already been suggested for the treatment of COVID-19, was also found to be effective as the inhibitor of RdRp of SARS- CoV-2[14]. Although their binding affinity was lower than the aforementioned top five drugs, the binding pattern was almost similar to these drugs.The drug surface hotspot study revealed that the molecular binding sites of all the five compounds were of a similar pattern that could be the hotspot of drug binding to the RdRp of SARS-CoV-2. This study suggested that amino acids Y32, K47, Y122, Y129, H133, N138,

D140, T141, S709 and N781 in RdRp of SARS-CoV-2 could make effective interactions with drugs, though this needs to be validated in wet lab. Additionally, pharmacophore was designed using the top RdRp inhibitor drugs along with Remdesivir, which was used further for the screening of ZINC database. Molecular docking analysis revealed that two compounds among the four hits had interacted effectively with the RdRp of SARS-CoV-2 which is indicativeof their potential as inhibitors of the enzyme. Although their binding affinity was lower than one control (Remdesivir), the vast number of interactions would give the complex its stability with this binding energy. Moreover, ADME and toxicity analysis of these compounds suggested that they could be used for the development of new drugs to treat COVID-19. However, the study of cytochromes P450 (CYP) isoforms inhibition concluded that there was a possibility that the suggested compounds could interact with CYP isoforms.

## 5. Conclusion

COVID-19 has created a disastrous global crisis affecting thousands of people every day, having already claimed thousands of lives, and severely hampered the global economy. The present study aimed to combat this global crisis by suggesting potential drug candidates for the treatment of COVID-19. Rifabutin, Rifapentine, Fidaxomicin, 7-methyl-guanosine-5'-triphosphate-5'-guanosine, Ivermectin and other screened novel compounds made the common drug surface hotspot in the RdRp of SARS-CoV-2, highly suggesting that they could be effective in the treatment of SARS-CoV-2.

**Tables**

Table 1: Top 5 RdRp Inhibitors with Binding Energy and Involved Amino Acid with Position

| Sl. No. | Drug Bank ID | Name | Binding Energy (kcal/mol) | No of Non-Covalent Interactions | Involved Amino Acids with positions |
|---|---|---|---|---|---|
| 1 | DB00615 | Rifabutin | -11.8 | 7 | Y32, K47, Y129, H133, N138, C139, T141, S709 |
| 2 | DB01201 | Rifapentine | -11.6 | 10 | V31, Y32, R33, K47, K121, Y122, Y129, H133, N138, C139, D140, T141, S709, N781 |
| 3 | DB08874 | Fidaxomicin | -10.9 | 16 | Y32, K47, Y129, N131, H133, N138, D140, T141, S709, T710, D711, K714, N781, Q773 |
| 4 | DB03958 | 7-methyl-guanosine-5'-triphosphate-5'-guanosine | -10 | 12 | Y32, R33, K47, Y122, Y129, H133, D140, T141, A706, S709, T710, D711, , G774, N781 |
| 5 | DB00602 | Ivermectin | -9.9 | 9 | Y32, L49, Y129, H133, S709, T710, K714, G774, N781 |
| Control | DB14761 | Remdesivir | -8.8 | 5 | K47, Y129, A130, H133, F134, D135, N138, C139, T141, S709, T710, D711, Q773 and N781 |
| Control | DB12466 | Favipiravir | -5.3 | 6 | Y129, H133, S709, K780 and N781 |

Table 2: Screened Compounds Molecular docking Result with involved amino acid

| Sl. No. | ZINC ID | Compound IUPAC Name | Binding Energy kcal/mol | No. of Non-covalent Interactions | Involved Amino Acids |
|---|---|---|---|---|---|
| 1 | ZINC09128258 | [(1,1-dioxo-1λ□-thiolan-3-yl)(2-methylpropyl)carbamoyl]methyl 3-(furan-2-amido)-4-methylbenzoate | -7.1 | 6 | Y32, K47, H133, D135, A706, S709, T710. D711, K714, G774, N781, S784 |
| 2 | ZINC09883305 | [(butan-2-yl)(1,1-dioxo-1λ□-thiolan-3-yl)carbamoyl]methyl 3-(furan-2-amido)-4-methylbenzoate | -7 | 9 | Y129, H133, D135, N138, A706, S709, T710, K780, N781, S784 |
| Control | Drug Bank ID: DB14761 | Drug Name: Remdesivir | -8.8 | 5 | K47, Y129, A130, H133, F134, D135, N138, C139, T141, S709, T710, D711, Q773 and N781 |
| Control | Drug Bank ID: DB12466 | Drug Name: Favipiravir | -5.3 | 6 | Y129, H133, S709, K780 and N781 |

Table 3: Drug Likeness Properties analysis of Screened compounds from ZINC 15 Database

| Drug Likeness Properties | ZINC09128258 | ZINC09883305 |
|---|---|---|
| Formula | C23H28N2O7S | C23H28N2O7S |
| Molecular Weight (g/mol) | 476.54 | 476.54 |
| Molar Refractivity | 122.38 | 122.38 |
| TPSA (Topological Polar Surface Area) | 131.37 | 131.37 |
| **Lipophilicity** | | |
| Log $P_{o/w}$ (iLOGP) | 3.15 | 3.01 |
| Log Po/w (XLOGP3) | 2.8 | 2.8 |
| Log Po/w (WLOGP) | 3.56 | 3.7 |
| Log Po/w (MLOGP) | 1.28 | 1.28 |
| Log Po/w (SILICOSNoIT) | 2.62 | 2.62 |
| Consensus Log Po/w | 2.68 | 2.68 |
| **Solubility** | | |
| LOG S (SILICOS-IT) | -5.77 | -5.77 |
| SILICOS-IT Solubility (mg/ml) | 8.02E-04 | 8.02E-04 |
| SILICOS-IT Solubility (mol/l) | 1.68E-06 | 1.68E-06 |
| Solubility class | Moderately soluble | Moderately soluble |
| **Pharmacokinetics** | | |
| Druglikeness | -3.39 | 1.84 |
| Drug-score | 0.34 | 0.6 |
| Blood-Brain-Barrier Permeant | No | No |
| Human Intestinal Absorption | Yes | Yes |
| Caco-2 Permeant | No | No |
| P-glycoprotein Substrate | No | Yes |
| CYP450 1A2 Inhibitor | No | No |
| CYP450 2C9 Inhibitor | Yes | Yes |
| CYP450 2D6 Inhibitor | Yes | Yes |
| CYP450 2C19 Inhibitor | Yes | Yes |
| CYP450 3A4 Inhibitor | Yes | Yes |
| CYP Inhibitory Promiscuity | Yes | Yes |
| **Toxicity** | | |
| AMES Toxicity | No | No |
| Carcinogens | No | No |
| Biodegradation | No | No |
| Acute Oral Toxicity (kg/mol) | III, 2.787 | III, 2.731 |
| Mutagenicity | No | No |
| Tumorigenicity | No | No |
| Irritating effects | No | No |
| Reproductive effects | No | No |

**Figure legends**

Fig. 1: Docked figures of Top 5 drug candidates with RdRp of SARS-CoV-2

Fig. 2: The interaction of RdRp with (a) Rifabutin (b) Rifapentine (c) Fidaxomicin (d) 7-methyl-guanosine-5'-triphosphate-5'-guanosine and (e) Ivermectin. Here, Drugs are in orange while protein active site pockets are in cyan lines. Solid blue lines represent H-bonds, while hydrophobic interactions are gray dashed lines. In addition, salt bridges, π-cation stacking, and halogen contacts are represented by yellow spheres connected by black dashed lines, orange dashed lines, and cyan lines, respectively.

Fig. 3: Structural Analysis of Drug Hotspot in RdRp of SARS-CoV-2

Fig. 4: Ligand-based pharmacophore model of RdRp of SARS-CoV-2. Here, green represents the hydrophobic features, violet represents aromatic features and yellow represents the hydrogen acceptor features.

Fig. 5: Docked figures of selected ZINC compounds with RdRp of SARS-CoV-2.

Fig. 6: The interaction of RdRp with (a) ZINC09128258 and (b) ZINC09883305. Here, Drugs are in orange while protein active site pockets are in cyan lines. Solid blue lines represent H-bonds, while hydrophobic interactions are gray dashed lines. In addition, salt bridges, π-cation stacking, and halogen contacts are represented by yellow spheres connected by black dashed lines, orange dashed lines, and cyan lines, respectively.

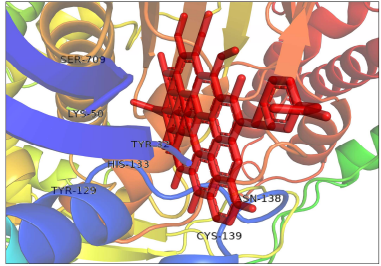
**Rifabutin (-11.8 kcal/mol)**

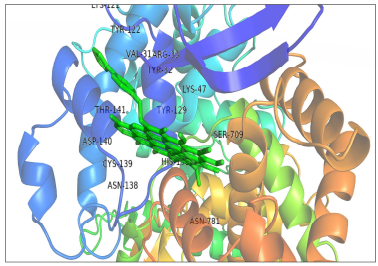
**Rifapentine (-11.6 kcal/mol)**

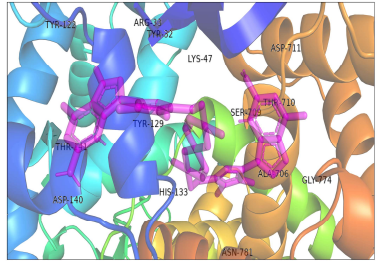
**Fidaxomicin (-10.9 kcal/mol)**

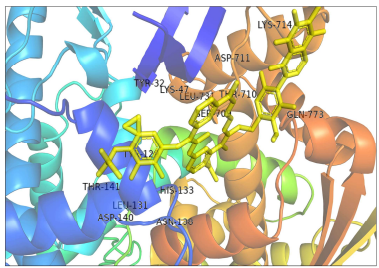
**7MGTPG (-10.0 kcal/mol)**

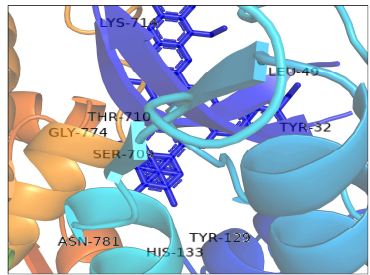
**Ivermectin (-9.9 kcal/mol)**

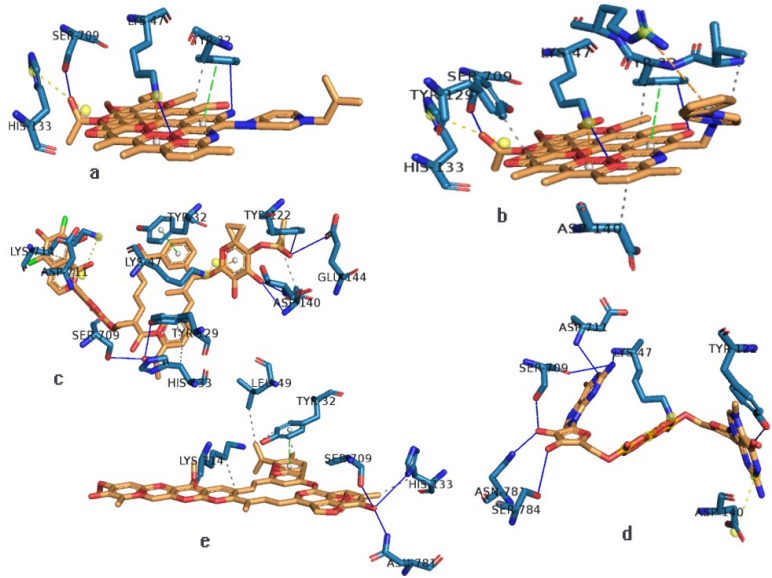

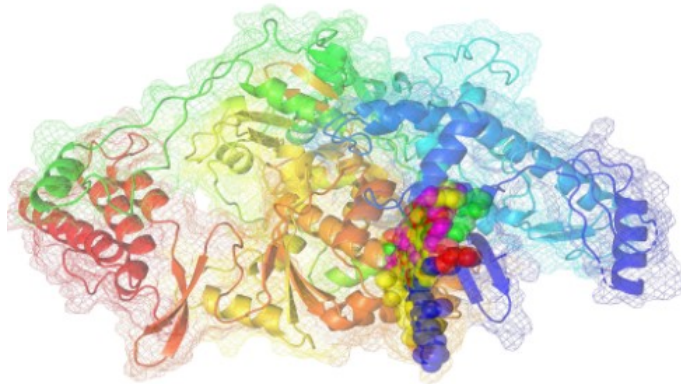

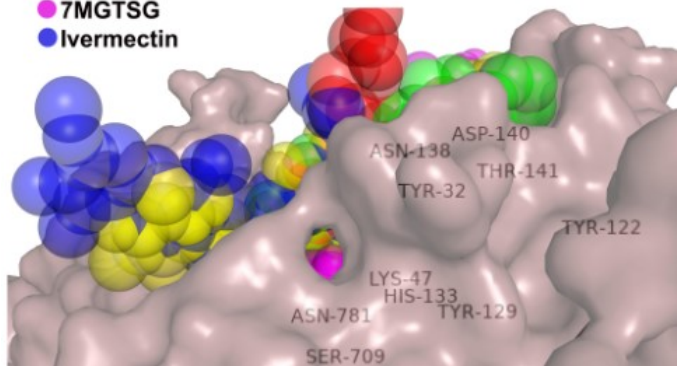

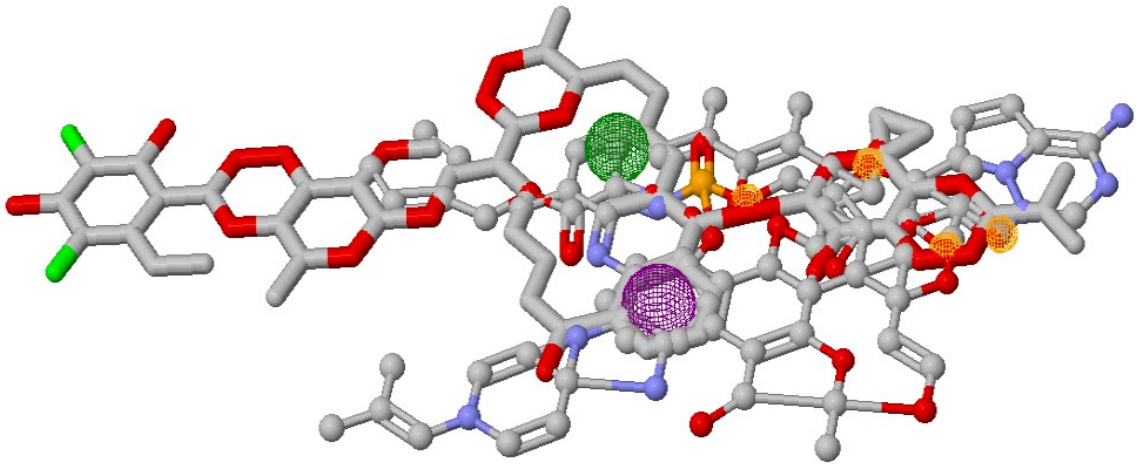

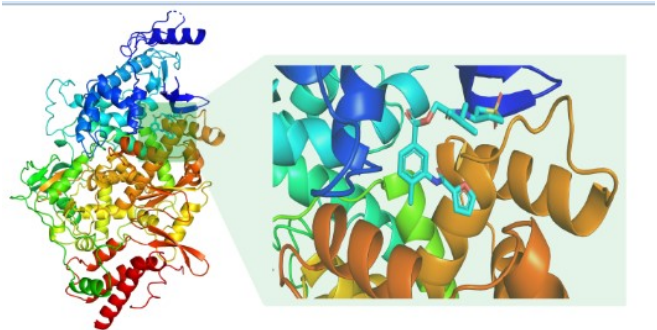

ZINC09128258

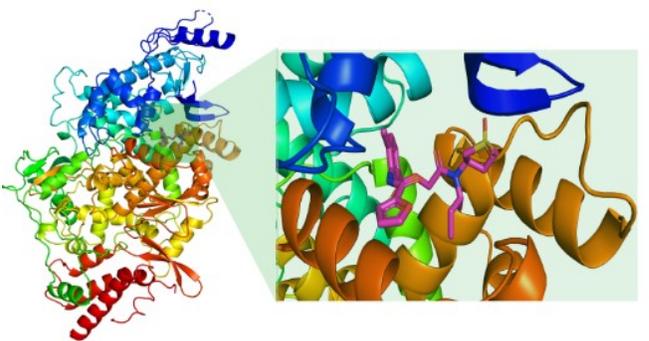

ZINC09883305

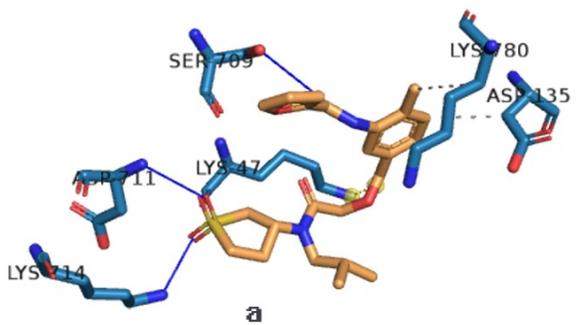

a

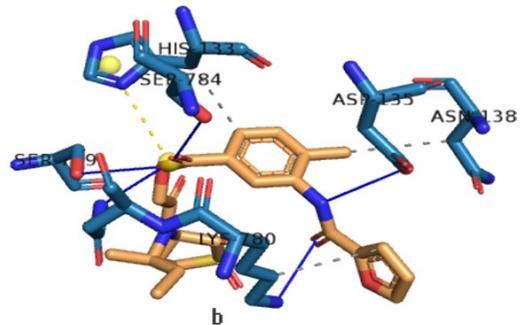

b